\documentstyle[12pt,dina4,psfig,preprint,aps]{revtex}

\begin{document}
\draft

\title{The Disappearance of Flow\footnote{supported by GSI, BMFT and DFG}}
\author{
        Sven Soff$^{a}$, Steffen A. Bass$^{a,b}$,
        Christoph Hartnack$^{b,c}$, Horst St\"ocker$^a$ and Walter
Greiner$^a$}
\address{
        $^a$Institut f\"ur Theoretische Physik der J. W.
Goethe-Universit\"at\\
        $^{~}$Postfach 11 19 32, D-60054 Frankfurt am Main, Germany\\
        $^b$GSI Darmstadt, Postfach 11 05 52, D-64220 Darmstadt, Germany\\
        $^c$\'Ecole des Mines, F-44072, Nantes, France
}

\maketitle

\begin{abstract}




We investigate the disappearance of collective flow in the reaction plane in
heavy-ion collisions within a microscopic model (QMD).\\
A systematic study of the impact parameter
dependence is performed for the system Ca+Ca.
The balance energy strongly increases with impact parameter.
Momentum dependent interactions
reduce the balance energies for intermediate
impact parameters $b\approx4.5$ fm. Dynamical negative
flow is not visible in the
laboratory frame but does exist in the contact
frame for the heavy system Au+Au.
For semi-peripheral collisions of Ca+Ca with
\mbox{$b\approx6.5$ fm} a new
two-component flow is discussed.
Azimuthal distributions exhibit strong collectiv flow signals,
even at the balance energy.
\end{abstract}


\newpage

\section{Introduction}

The prediction of collective flow in heavy-ion collisions
by the hydrodynamical model \cite{sch74} has yielded a powerful tool for the
investigation of excited nuclear matter.
Main goals are to determine the equation of state ({\em eos})
and the in-medium
nucleon-nucleon cross section.
One possible approach is the measurement
and calculation of the transverse flow in
the reaction plane.
At beam energies above $100-200$ AMeV
two-body collisions rule the dynamics
yielding the typical bounce-off behaviour \cite{st80,do86,st86,va88},
which is the deflection of
cold spectator matter from hot compressed participant matter.
The attractive part of the mean field becomes
more and more important with a decrease in energy.
As a consequence even negative scattering angles are
possible \cite{moli85a} which can be imagined as partial orbiting of
the two nuclei \cite{we93}.
At a certain incident energy, called the balance
energy $E_{bal}$, the attractive and repulsive forces which are
responsible for
the transverse flow in the reaction plane cancel each other,
causing the disappearance of this particular flow characteristic. \\
The notation ``energy of vanishing flow'',
as the balance energy is often called, can lead to misunderstandings:
In particular, we will demonstrate by inspecting azimuthal distributions
that strong flow still exists
at the balance energy.
Whereas it was shown for small  impact parameters
that the balance energy depends only
weakly on the stiffness of the equation of state \cite{xu91,ha93},
a large sensitivity to the nucleon-nucleon
in-medium cross section was recognized \cite{xu91,ha93,we93}.
The functional dependence of the balance energy
on the system size can be approximately described by a power law:
$E_{bal} \sim A_{tot}^{-\frac{1}{3}}$ \cite{mota92,we93}.
Systematic studies of the mass dependence of the disappearance
of flow proposed a reduction of the in-medium cross section of about
$20\%$ with respect to the free NN-cross section at normal
nuclear density\cite{we93} by comparing the
measured data \cite{og90,zhang90,su90,krof91,krof92,pet92,we93}
with BUU calculations.
However, all investigations neglected to study the impact parameter dependence
of the disappearance of flow.\\
In this contribution we show that a variation of the impact parameter
changes decisively the balance energy $E_{bal}(b)$ and as a consequence
the mass dependence analysis receives an important new variable.\\
The system Au+Au exhibits no negative flow in the laboratory frame.
However, if the initial pre-contact rotation of the system due to
Rutherford-trajectories is subtracted, large negative flow appears.\\
A new two-component flow appears in collisions with large impact
parameters.\\
Azimuthal asymmetries persist at
the balance energy.\\
The balance energy $E_{bal}$ is nearly independent of particle
type \cite{we93}, although it is well
known that the strength of the flow depends on it.
Therefore we will mostly regard all nucleons
and check the effect of taking clustering
into account.

\section{The model}

The {\bf Q}uantum {\bf M}olecular
{\bf D}ynamics model (QMD)\cite{ha93,ai86,ai87b,ha88,ha89,pei89,ai91}
is employed here.
In the QMD model the nucleons are represented by Gaussian
shaped density distributions. They are initialized in a sphere of a radius
$R=1.12A^{1/3}$ fm, according to the liquid drop model. Each nucleon is
supposed to occupy a volume of $h^3$, so that the phase space is uniformly
filled. The initial momenta are randomly chosen between zero and the local
Thomas-Fermi-momentum. The $A_P$ and $A_T$ nucleons interact
via two- and three-body
Skyrme forces, a Yukawa potential, momentum dependent interactions, a
symmetry potential (to achieve a correct distribution of protons and neutrons
in the nucleus) and explicit Coulomb forces between the $Z_P$ and $Z_T$
protons.
Using this ansatz we have chosen a hard equation of state
with a compressibility of $\kappa=380$ MeV
\cite{moli85b,kru85a}.
For the momentum dependent interaction we use a phenomenological ansatz
\cite{schue87,ai87b,bert88b}
which fits experimental measurements \cite{ar82,pa67}
of the real part of the nucleon optical potential.
The nucleons  are propagated according to Hamiltons equations of motion.
A clear distinction is made between protons
and neutrons with Coulomb forces acting only on the protons and an
asymmetry potential containing the asymmetry term from the
Bethe--Weizs\"acker
formula acting between protons and neutrons. Furthermore parameterized
energy dependent free
$pn$ and $pp$ cross sections are used instead of an averaged nucleon--nucleon
cross section. They differ by 50\% at 150 MeV. It was shown that their
energy dependence cannot be neglected \cite{li93}.
Hard N-N-collisions are included by employing
the collision term of the well known VUU/BUU
equation \cite{st86,kru85a,bert84,ai85a,wo90,lib91a}. The collisions are done
stochastically, in a similar way as in the CASCADE models \cite{yar79,cug80}.
In addition, the Pauli blocking (for the final state) is taken into account
by regarding the phase space densities in the final states of a two body
collision.
\section{Results and Discussion}
For the investigation of transverse
flow in the reaction-plane the in-plane
transverse momentum $p_{x}$ is usually plotted versus the normalized
rapidity $y/y_{p}$.
Fig.1 shows
the $p_{x}(y)$ distribution at two different energies for the system
Ca+Ca and \mbox{$b=0.5b_{max}\approx4$ fm.}
At $80$ AMeV a negative slope (corresponding to negative scattering angles)
is visible whereas for $130$ AMeV
the opposite sign (positive scattering angles) is found.
The first corresponds to negative scattering angles of the
majority of the protons, the latter illustrates the deflection of
nucleons caused by nucleon-nucleon collisions.\\
In order to determine the balance energy, the energy
is varied between these two values
and a linear fit is applied to the slopes
of the $p_{x}(y)$ distributions.
These slopes, which are called reduced flow,
have negative values for energies
smaller than $E_{bal}$ and positive values for energies higher
than $E_{bal}$.
The balance energy $E_{bal}$ is obtained again by a
linear fit to the energy dependence of the reduced flow at the point where
the reduced flow passes through zero (fig.2).
Onethousand events of Ca+Ca are performed for
a hard equation of state without momentum dependent
interactions. Different symbols correspond to the
different impact parameters $0.25b_{max},0.4b_{max},0.5b_{max},
0.6b_{max}$.
The balance
energies differ completely for the different impact parameters. This is
in contrast
to claims in \cite{zhou93}.
The errors of the balance energies are approximately
$\pm5$ AMeV .\\
Fig.3 depicts
the impact parameter dependence of the balance energy for the system Ca+Ca.
An approximate linear increase of the
balance energy with impact parameter is visible. At larger impact
parameters fewer nucleon-nucleon collisions yield reduced repulsive
forces, therefore the attractive mean field dominates.
For larger impact parameters the balance energy
is smaller if momentum dependent interactions (mdi)
are included, due to their repulsive effects.
The balance energy is insensitive to
the inclusion of mdi for small impact parameters $b\le0.25b_{max}$.
The balance energy for Ca+Ca varies from $65$ to $150$ AMeV without mdi and
from $75$ to $115$ AMeV with mdi, depending on impact parameter.
Experiments \cite{we93} show the balance energy for Ar+Sc
, i.e. $ A=85$,
to be $87 \pm 12$ AMeV, the impact parameter was estimated
to be approximately $0.4b_{max}\approx3$fm.
This value is compatible with ours.
Even for rather central collisions
with a maximum impact parameter of $0.4b_{max}$
the balance energies for Ca+Ca reach values from $65$ AMeV up to
$95$ AMeV
depending on impact parameter.
This is a significant variation contrary to the claims in \cite{zhou93}.
A precise
knowledge of the impact parameter
is of utmost importance before any conclusions about the balance energy
concern the equation of state or the in-medium nucleon-nucleon cross
section.\\
Let us now turn to a different question:
Two flow-components appear in one event
showing both positive and negative
flow if semi-peripheral collisions of Ca+Ca at $b=0.85b_{max}\approx6.5$ fm
and $E=350$ AMeV are considered.
Fig.4 illustrates this effect.
The nucleons show positive $p_{x}$-values for small rapidities on the
average in the forward hemisphere ($y_{cm}\ge0$) whereas negative
$p_{x}$-values
are observed for higher rapidities.
This effect is seen for the hard equation of state without momentum
dependent interactions, it is very sensitive
to the incident energy,
the impact parameter, and most importantly,
to the addition of momentum dependent interactions (mdi).\\
The signs of the average
$p_{x}$-values become positive for all positive rapidities if
the impact parameter is reduced to $b=0.7b_{max}$.
The same happens if momentum dependent interactions
(which give additional repulsion) are introduced.
The following scenario might explain the two components: Nucleons
which have experienced
higher densities, e.g. $\rho_{max}\ge1.3\rho_{0}$ are preferentially
visible at small rapidities.
\hyphenation{posi-tive}
This compressed, stopped matter shows positive flow.
The spectator matter,
which has experienced less compression, shows negative flow.
The separation of the two components is clearly visible when applying a cut
on the maximum density for slightly different system parameters
($E=330$ AMeV and $b=0.75b_{max}$).
In addition the components can be separated with respect
to their type in a simple configuration space coalescence model \cite{pei89}.
Protons yield the major
part of the component at midrapidity whereas heavier fragments
rule the outer component.
The time-evolution of the collision can be imagined as if
the spectators were sucked to the participant zone.\\
A two-component flow is observed
for Ca+Ca at $170$ AMeV e.g. at $b=0.8b_{max}$
with momentum dependent interactions, too.
The sign of the components at midrapidity and larger rapidities
is just opposite to those observed without mdi.
Coalescence considerations indicate in turn that the components
around midrapidity stem from heavier fragments, while
free nucleons
contribute mainly at $y/y_{proj.}\ge \pm1$.\\
This double sign change is highly sensitive to
the momentum dependent interactions and should therefore be
experimentally scrutinized. \\
Let us now turn to another point:
A smaller balance energy $E_{bal}$ is expected for the heavy system Au+Au
($A=394$) than for Ca+Ca due to the
cited $A^{-\frac{1}{3}}$-law. Experimentally so far only an upper bound
for the balance energy of $E_{bal} \le 60$ AMeV \cite{zhang90} has
been found.
Therefore the existence of negative flow
is an open question due to the strong Coulomb repulsion.
We show that this is due to an ill-defined frame of reference.
The flow is in fact balanced at
$E_{bal}=55 \pm 5$ AMeV and $E_{bal}=65 \pm 5$
AMeV for the
impact parameters $b=0.25b_{max} \approx 3.3$ fm and $b=0.5b_{max} \approx
 6.5$ fm, respectively and for a hard equation of state without momentum
dependent interactions.
These values are obtained if the initial pre-contact rotation of the
system due to Rutherford-trajectories
is subtracted. In this system the sign-reversal for the
reduced flow is clearly visible. Fig.5 shows the respective
calculation for Au+Au at $50$ AMeV and $b=0.5b_{max}\approx6.5$ fm.
In the rotated system the flow is obviously negative whereas a
flat distribution is obtained in the laboratory frame.
In the laboratory frame negative flow does not appear
for any impact parameter, even not for
low energies.\\
Let us now turn to  the squeeze-out which is an
established effect \cite{st82,gut89,dem90}. Excited
participant matter is pushed out perpendicular to the reaction plane.
At energies dicussed in this paper this behaviour might be different.
In fig.6 these azimuthal angular-distributions are
plotted for the system Ca+Ca (hard eos+mdi)
at their respective balance energies with different impact
parameters.
The considered rapidity is $-0.15 \le y/y_{p} \le 0.15$
according to recent experiments for the heavier system Zn+Ni \cite{pop94}.
The full lines are the result of fits by the Legendre-expansion:
$dN/d\phi = a_0(1+a_1cos(\phi)+a_2cos(2\phi))$.
The value of $a_2$ gives a measure of the anisotropy of this
collective motion. Negative values of $a_2$ show
prefered emission perpendicular to the reaction plane whereas
positive values describe an enhancement in the reaction-plane.
Fig.6 shows that for Ca+Ca the in-plane emission is
prefered for larger impact parameters,
and a slight out-of-plane enhancement is observed
for rather central collisions at the balance energies and at midrapidity.
The transition energy where the anisotropy parameter $a_2$
becomes zero, corresponding to an azimuthally symmetrical distribution,
was measured for Zn+Ni \cite{pop94}. It was found that this
transition energy is smaller than the corresponding balance energy.
Our calculations for the lighter system Ca+Ca show the transition
energies to be larger than the balance energy for
larger impact parameter ($b\geq0.4b_{max}$),
but smaller for more central collisions.
This was already indicated by measurements for Ar+V \cite{wi90}.
Measurements indicate that the
in-plane enhancement increases with impact parameter \cite{sh93}.
This can be seen in fig.7
for Ca+Ca at $80$ AMeV and various
impact parameters.
Light fragments show a slightly more pronounced in-plane to out-of-plane ratio
than single nucleons if clustering is taken into account.
Consequently, it must be pointed out that even at the in-plane
balance energy collective flow characteristics are clearly visible in the
azimuthal distributions.
\section{Conclusions}
We have investigated the disappearance of the in-plane flow
for Ca+Ca and Au+Au.\\
$\bullet$A strong impact parameter dependence of
the in-plane balance energy $E_{bal}$ is observed.
The balance energy clearly increases with impact parameter.
This cannot be neglected
while pinning
down basic properties of
excited nuclear matter.\\
$\bullet$The balance energy is smaller with momentum dependent
interactions than without for large impact parameters.
The difference might be a tool to get information about the
proper parametrization of the momentum dependent interactions.\\
$\bullet$Negative
flow angles will not
be visible in the laboratory frame for the heavy Au+Au system
due to the long range Coulomb forces, although
the in-plane flow disappears. Negative flow and the respective
balance energies are visible in the frame where the pre-contact rotation due
to
the initial Rutherford-trajectories is subtracted.
However, a maximum mass must exist where negative
flow can still be observed in the laboratory frame.\\
$\bullet$A new two-component flow was shown for
large impact parameters. One component
stems from participant particles
at rapidities around $y_{cm}$ whereas the
other component results from cold spectator matter.
They show opposite sign in the $p_x(y)$-distribution.
The existence of two distinctly different
flow-components depends on
the inclusion of momentum dependent
interactions. This is of great importance for the proper
determination of the parametrization of
the momentum dependent interactions or other
basic properties such as the in-medium NN-cross
section.\\
$\bullet$Finally, azimuthal distributions
demonstrate the existence of flow, even at the
balance energy. For the system Ca+Ca the
energy of the change from an preferentially in-plane to
out-of-plane emission is smaller
for central collisions and larger for increasing
impact parameters than the balance energy.
This energy of an azimuthally symmetrical distribution can
provide valuable information complementary to the in-plane balance
energy. The in-plane to out-of-plane ratio increases with impact parameter.\\
The search for tools to describe excited nuclear matter in nucleus-nucleus
collisions and the search for signals to
determine unambigiously the basic physical
attributes is going on.

\pagebreak

\newpage
\pagestyle{empty}
\begin{figure}[h]
\vspace{-1.9cm}
\centerline{\psfig{figure=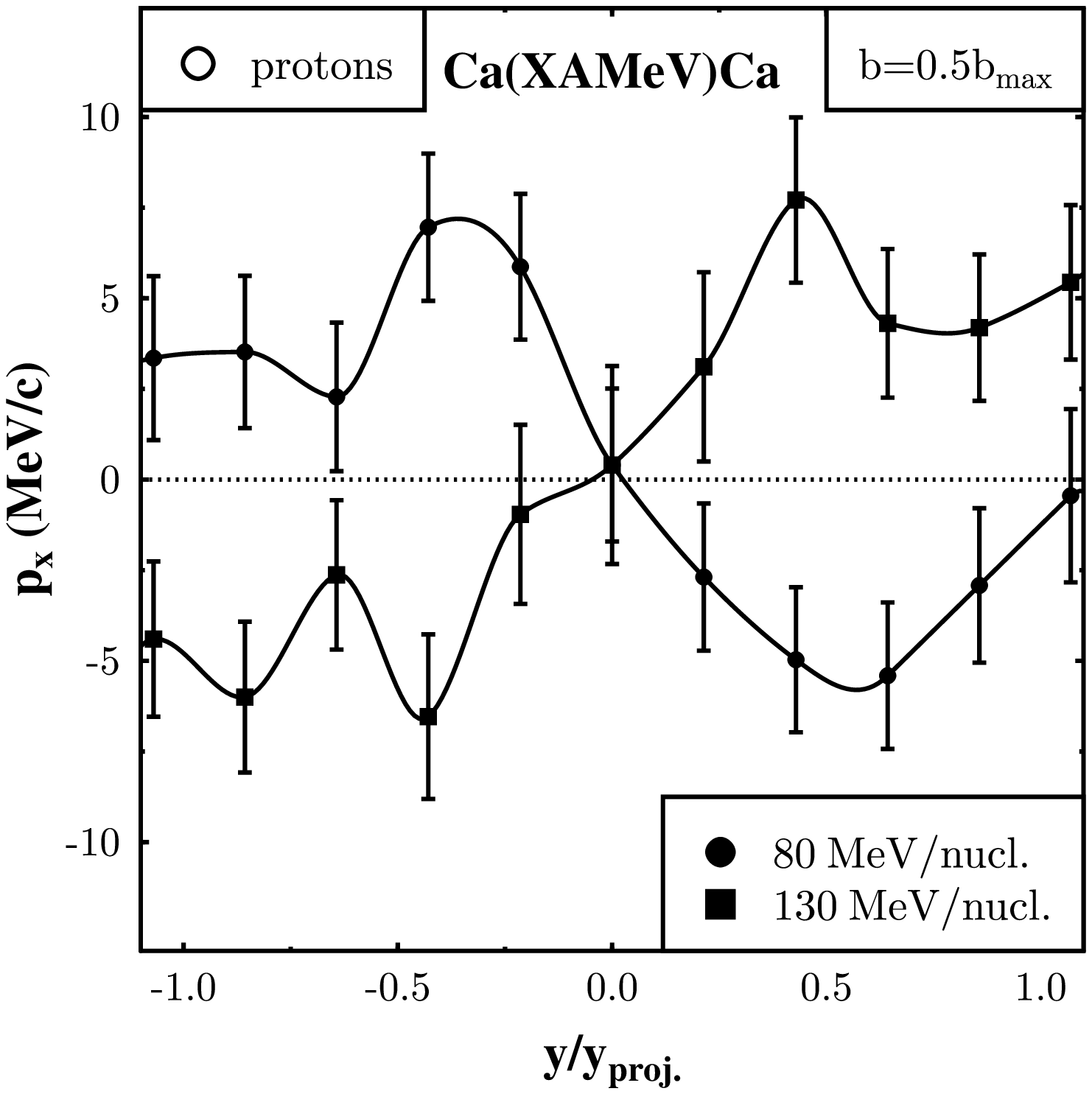}}
\vspace{-0.7cm}
\caption{Transverse momentum projected onto the reaction plane $p_x$
as a function of the normalized rapidity. This $p_x(y/y_p)$-distribution
of protons for the system Ca+Ca is plotted
at the two incident energies, $80$ AMeV and $130$ AMeV.
The impact parameter is half the maximum impact parameter $b=0.5 b_{max}$.
For each curve thousand events were calculated
with a hard equation of state without
momentum dependent interactions. The lines are plotted to guide the eye.}
\end{figure}

\newpage

\begin{figure}[h]
\vspace{-5.3cm}
\centerline{\psfig{figure=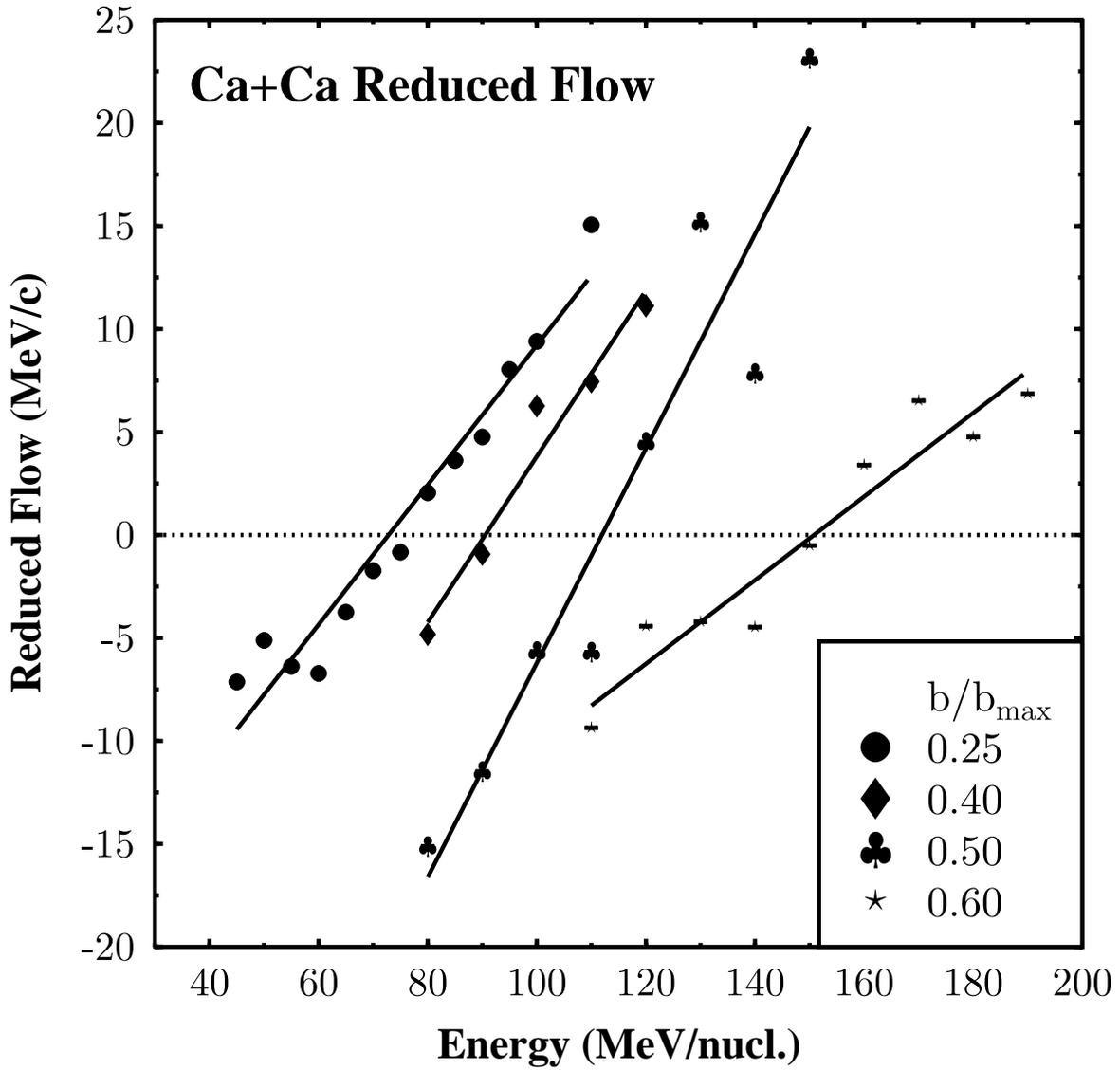}}
\vspace{-0.7cm}
\caption{Reduced flow values as a function of incident energy
and impact parameter for Ca+Ca. The impact parameters are $0.25, 0.4, 0.5,
0.6$ times the maximum impact parameter. Each point is a result of
thousand events with a hard equation of state without momentum dependence.
The straight lines are the results of linear fits.}
\end{figure}

\newpage

\begin{figure}[h]
\vspace{-5.3cm}
\centerline{\psfig{figure=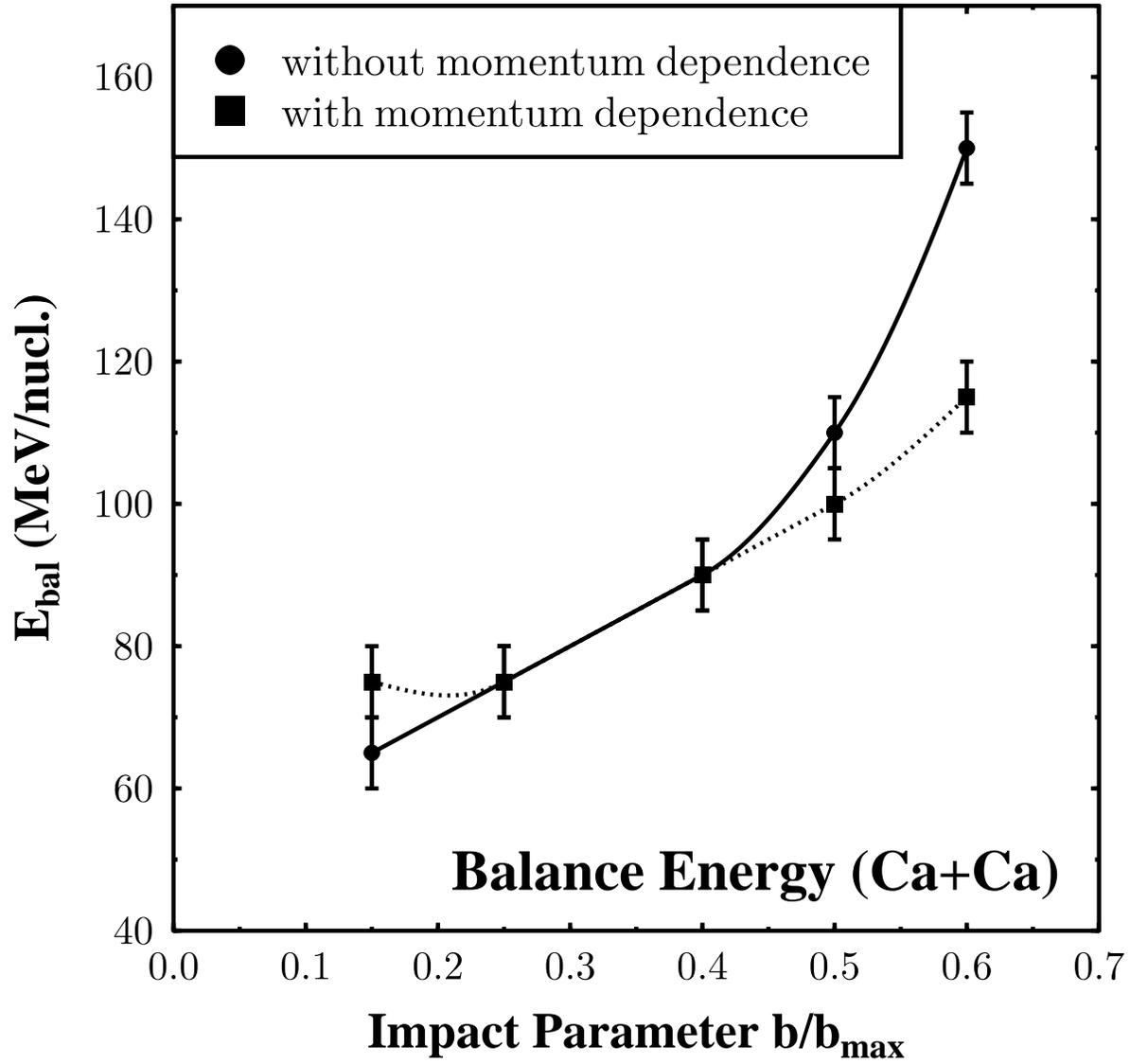}}
\caption{The in-plane balance energy $E_{bal}$
as a function of impact parameter $b$ for the system Ca+Ca.
The circels and squares are the calculated values without and with
momentum dependent interactions, respectively. The curves are plotted to guide
the eye.}
\end{figure}

\newpage

\begin{figure}[h]
\vspace{-5.3cm}
\centerline{\psfig{figure=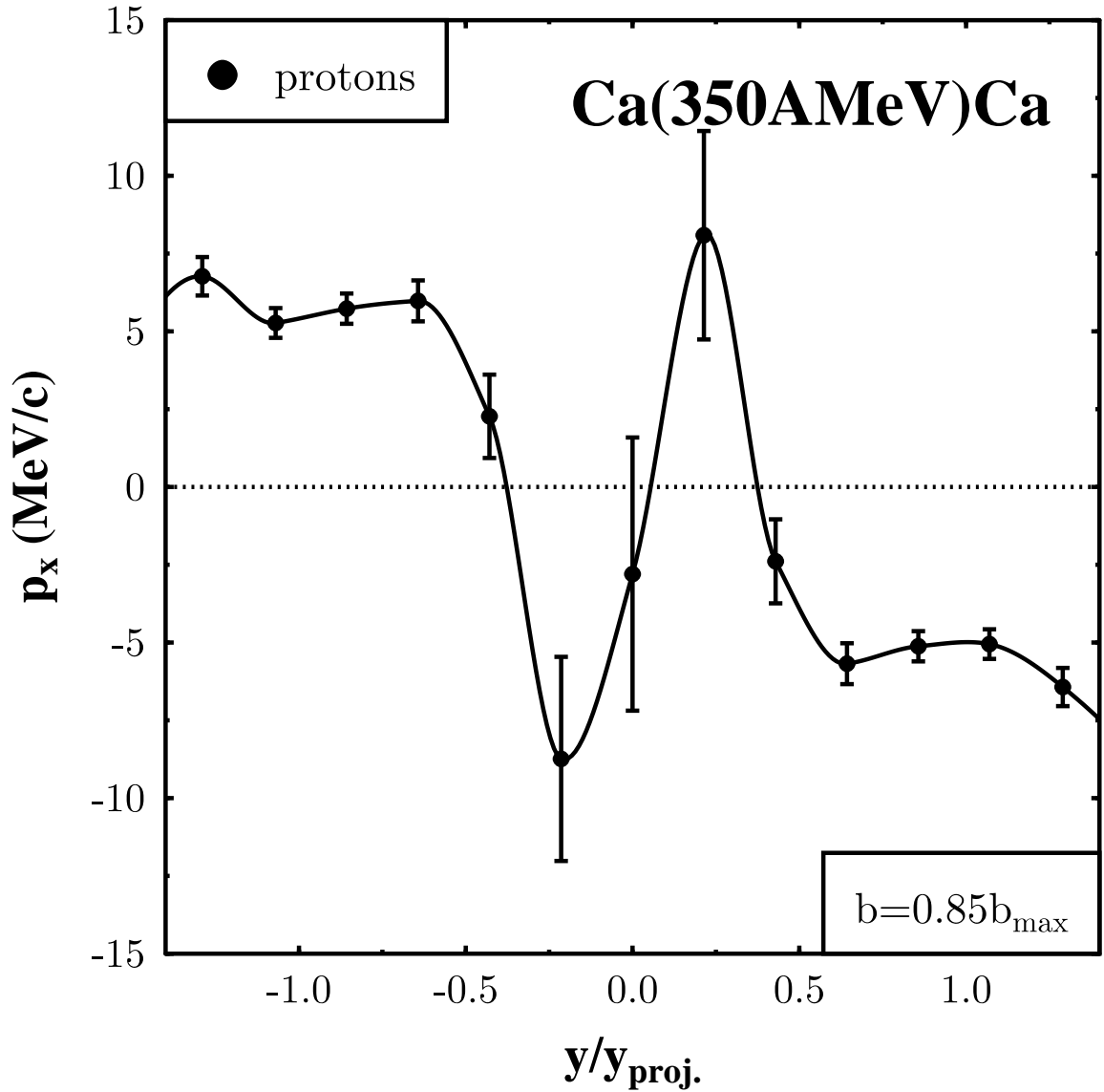}}
\caption{$p_x(y/y_p)$-distribution of protons for the semi-peripheral
($b=0.85b_{max}$) collision of Ca+Ca at $350$ AMeV incident energy.
This two-component flow is received by a calculation of 10000 events
with a hard equation of state without momentum dependent interactions.}
\end{figure}

\newpage


\newpage

\begin{figure}[h]
\vspace{-5.3cm}
\centerline{\psfig{figure=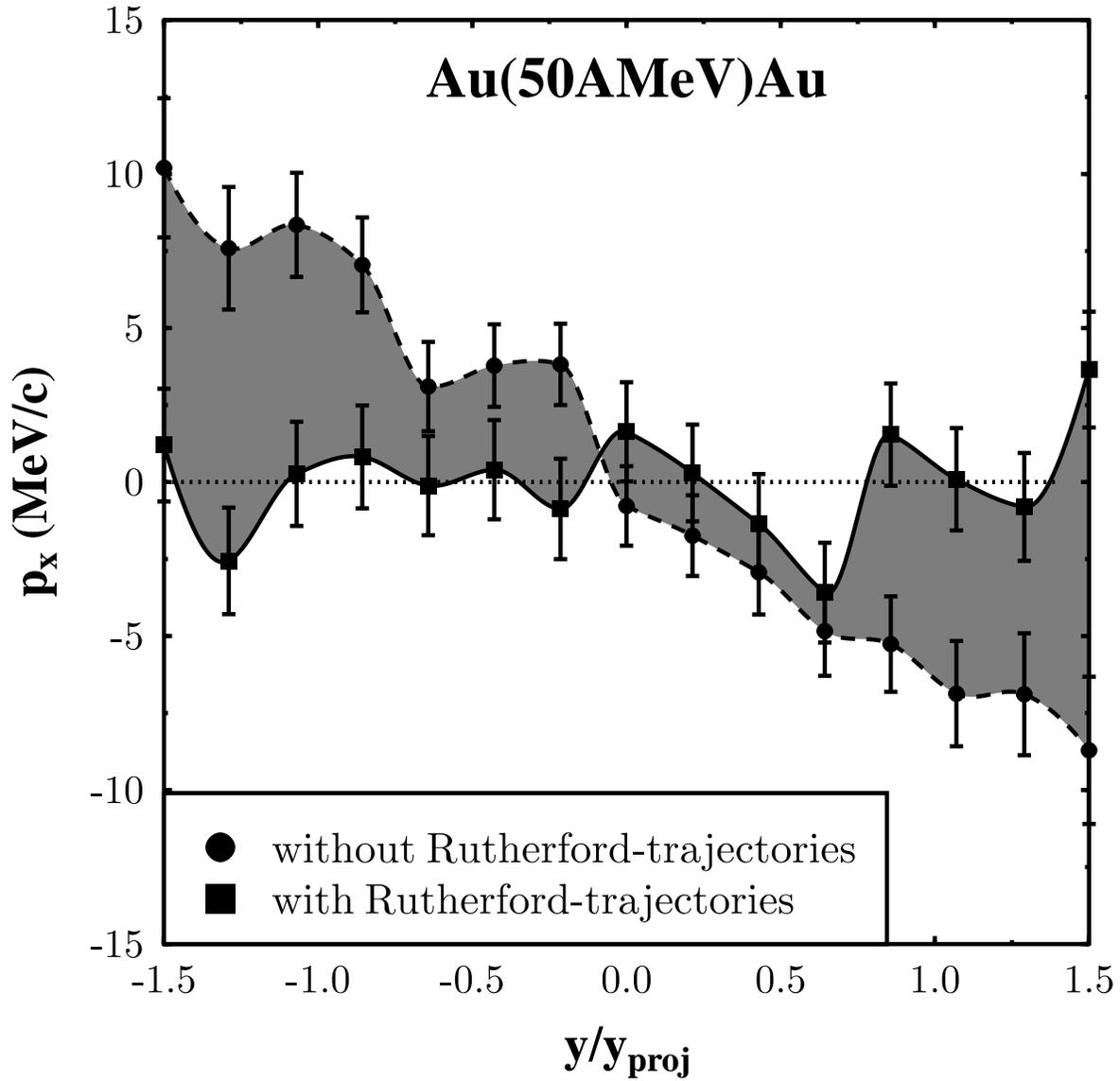}}
\vspace{-0.7cm}
\caption{$p_x(y/y_p)$-distribution of protons for the system Au+Au at
$50$ AMeV. The impact parameter is $0.5b_{max}$. The squares and circles
correspond to calculations with and without an initialization on
Rutherford trajectories. 500 events were calculated for each curve with
a hard equation of state
without momentum dependence.}
\end{figure}

\newpage
\samepage{
\begin{figure}[h]
\centerline{\psfig{figure=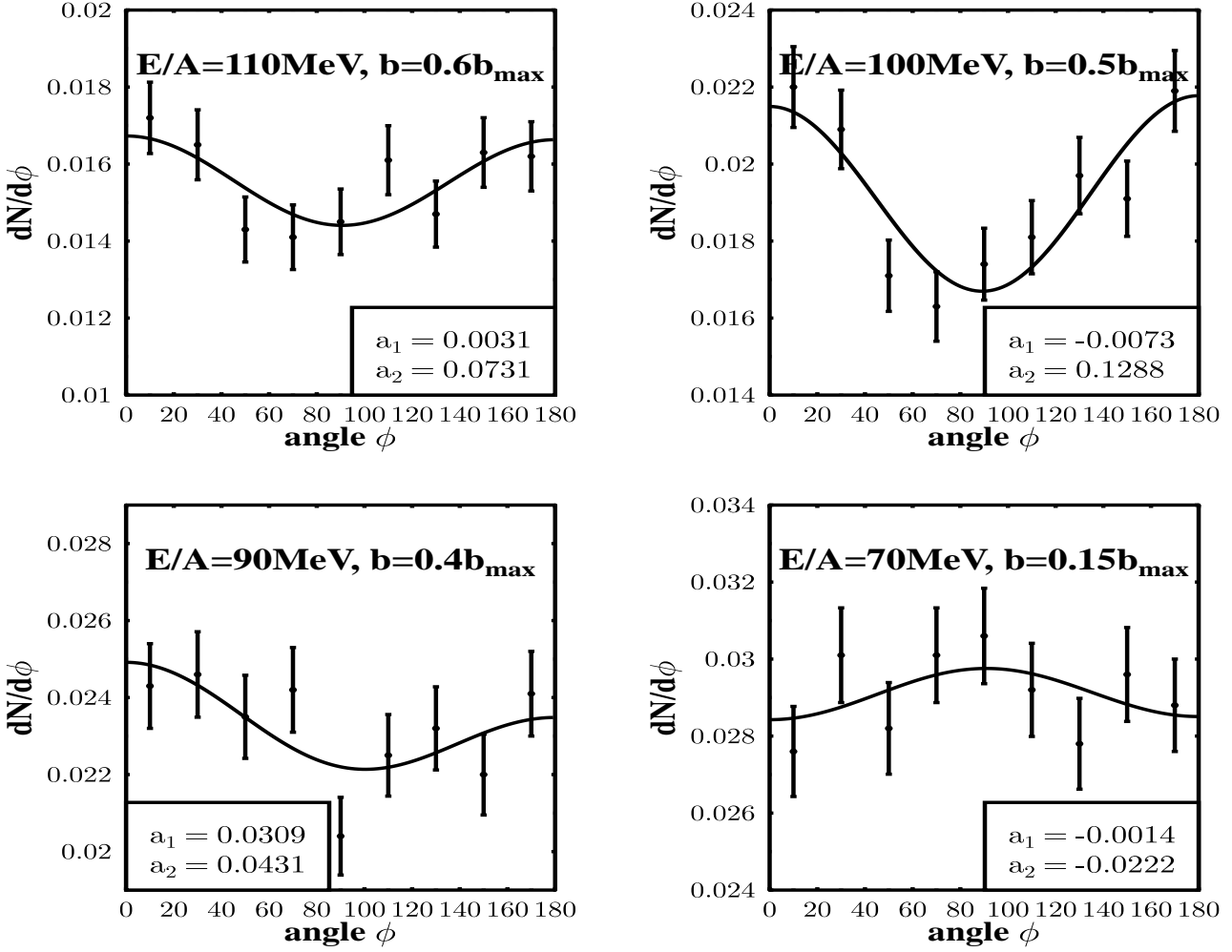,height=20cm,width=18cm}}
\vspace{-4.0cm}
\caption{Azimuthal distributions with respect to the reaction plane
for Ca+Ca. The incident energies and impact parameters correspond
to the determined in-plane balance energies $E_{bal}(b)$ with
momentum dependent interactions. The rapidity range is restricted to
$-0.15 \le y/y_p \le 0.15$. The curves are fits according to
$dN/d\phi=a_0(1+a_1cos(\phi)+a_2cos(2\phi))$.}
\end{figure}}

\newpage
\samepage{
\begin{figure}[h]
\vspace{-4.3cm}
\centerline{\psfig{figure=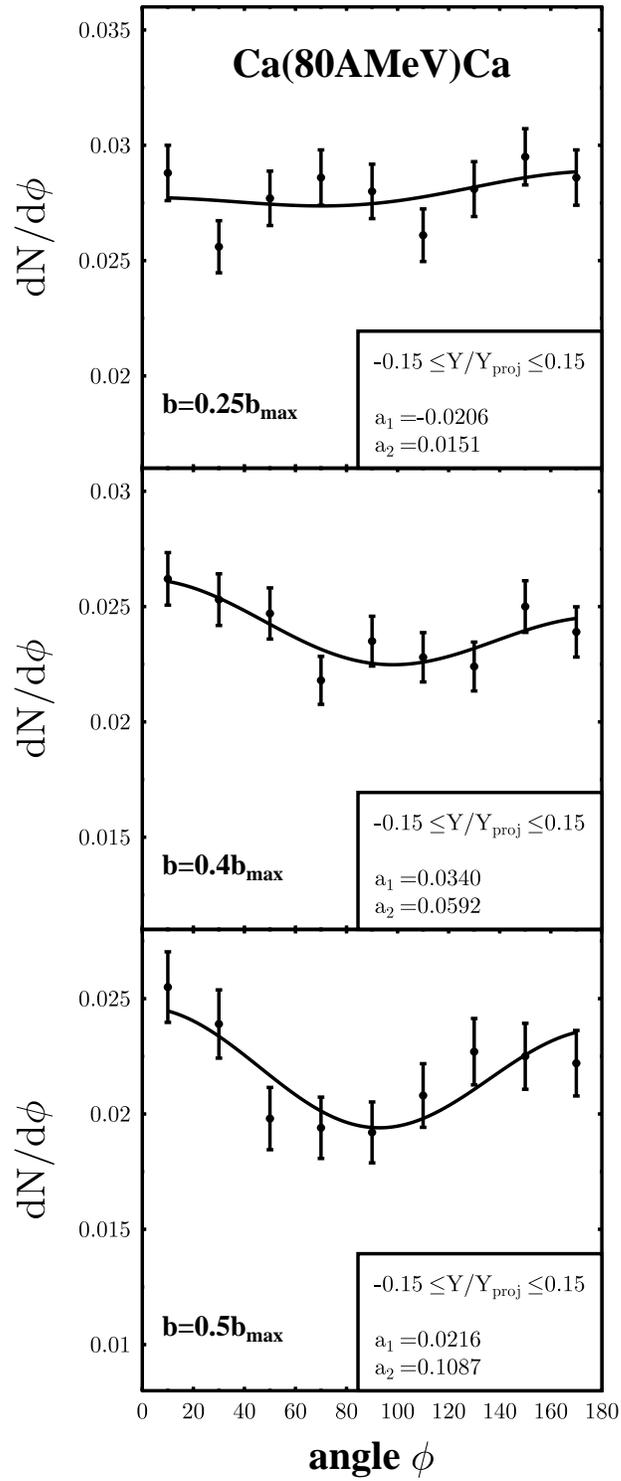,height=20cm}}
\nopagebreak
\vspace{-1.5cm}
\caption{Azimuthal distributions with respect to the reaction plane
for Ca+Ca at $80$ AMeV and for three different impact parameters
$b=0.25, 0.4,$ and $0.5 b_{max}$.}
\end{figure}}

\enddocument